# Room-temperature paramagnetoelectric effect in magnetoelectric multiferroics Pb(Fe$_{1/2}$Nb$_{1/2}$)O$_3$ and its solid solution with PbTiO$_3$

V. V. Laguta[1,3*], A. N. Morozovska[2†], E. A. Eliseev[3], I. P. Raevski[4], S. I. Raevskaya[4], E. I. Sitalo[4], S. A. Prosandeev[4,5], and L. Bellaiche[5].

[1]*Institute of Physics AS CR, Cukrovarnicka 10, 162 53 Prague, Czech Republic*

[2]*Institute of Physics, National Academy of Sciences of Ukraine, 46, pr. Nauky, 03028 Kyiv, Ukraine*

[3]*Institute for Problems of Materials Science, National Academy of Sciences of Ukraine, Krjijanovskogo 3, 03142 Kyiv, Ukraine*

[4]*Research Institute of Physics and Physical Faculty, Southern Federal University, Stachki Ave. 194, 344090, Rostov-on-Don, Russia.*

[5]*Physics Department and Institute for Nanoscience and Engineering, University of Arkansas, Fayetteville, Arkansas 72701, USA*

**Abstract**

We have observed the magnetoelectric response at room temperature and above in high-resistive ceramics made of multiferroic Pb(Fe$_{1/2}$Nb$_{1/2}$)O$_3$ (PFN) and PFN-based solid solution 0.91PFN – 0.09PbTiO$_3$ (PFN−PT). The value of the paramagnetoelectric (PME) coefficient shows a pronounced maximum near the ferroelectric-to-paraelectric phase transition temperature, $T_C$, and then decreases sharply to zero for $T > T_C$. The maximal PME coefficient in PFN is about $4\times10^{-18}$ s/A. The theoretical description of the PME effect, within the framework of a Landau theory of phase transitions allowing for realistic temperature dependences of spontaneous polarization, dielectric and magnetic susceptibilities, qualitatively reproduces well the temperature dependence of the PME coefficient. In particular, the Landau theory predicts the significant increase of the PME effect at low temperatures and near the temperature of the paraelectric-to-ferroelectric phase transition, since the PME coefficient is equal to the product of the spontaneous polarization, dielectric permittivity, square of magnetic susceptibility and the coefficient quantifying the strength of the biquadratic magnetoelectric coupling.

**Keywords**: magnetoelectrics, phase transitions, magnetoelectric coupling, phenomenological Landau theory, atomistic simulation

---

* Corresponding author 1: laguta@fzu.cz

† Corresponding author 2: anna.n.morozovska@gmail.com

# I. Introduction

Multiferroics are materials having two or more order parameters (for instance, magnetic, electric or elastic) coexisting in the same phase. They have emerged as an important topic in condensed matter physics due to both their intriguing physical behaviors and a broad variety of novel physical applications they enable. Among multiferroics, magnetoelectric (ME) materials, which exhibit coupling of electric polarization and magnetization are very promising for spintronic and magnetic random access memory applications. The unique physical properties of multiferroics originate from the complex interactions among the structural, polar and magnetic long-range order parameters [1]. For instance, biquadratic or "magnetocapacitive" and linear magnetoelectric couplings lead to intriguing effects, such as a giant magnetoelectric tunability of multiferroics [2]. Biquadratic coupling of the structural and polar order parameters, introduced in Refs. [3, 4, 5], are responsible for the unusual behavior of the dielectric and polar properties in ferroelastics – quantum paraelectrics.

In magnetoelectric materials, besides of the aforementioned linear and biquadratic couplings of magnetic and electric order parameters, linear-quadratic paramagnetoelectric (PME) effect should exist in the paramagnetic phase below the $T_C$ temperature of the paraelectric-to-ferroelectric phase transition, where the electric polarization is non-zero. This effect was first observed in piezoelectric paramagnetic crystal $NiSO_4 \cdot 6H_2O$ [6]. It was later measured in Mn-doped $SrTiO_3$ [7], magnetoelectric multiferroic $Pb(Fe_{1/2}Nb_{1/2})O_3$ (PFN) [8, 9, 10], and metal-organic framework [11]. Unexpectedly, it was found that the PME effect vanishes above the Neel temperature $T_N \approx 153$ K in PFN [10] while the $T_C$ of this system is about 360-370 K.

Within a phenomenological approach used in the work, linear and biquadratic ME couplings contribution to the system free energy are described by the terms $\mu_{ij} P_i M_j$ and $\xi_{ijkl} P_i P_j M_k M_l$, respectively (**P** is polarization and **M** is magnetization, and $\mu_{ij}$ and $\xi_{ijkl}$ are corresponding tensors of ME effects, respectively) [12, 13, 14, 15, 16]. The PME coupling contribution is described by the term $\lambda_{ijk} P_i M_j M_k$ [12, 13]. Sometimes it can be more convenient to use the E-H representative of the ME coupling terms, i.e., $\alpha_{ij} E_i H_j$, $\eta_{ijkl} E_i E_j H_k H_l$ and $\beta_{ijk} E_i H_j H_k$, respectively [12] ($E_i$ and $H_i$ are components of the electric and magnetic fields, respectively).

In general, the quadratic ME coupling is much less studied in magnetoelectrics than the linear coupling while the former one can be even larger than its linear counterpart in antiferromagnetic (AFM) materials, where only a weak magnetization exists due to possible canting of AFM ordered magnetic moments caused by the antisymmetric Dzyaloshinskii-Moriya interaction $\hat{H}_{DM} = \mathbf{D} \cdot [\mathbf{S_1} \times \mathbf{S_2}]$ [17].

In the present paper, we study in detail the temperature dependence of the PME effect of high-resistive PFN and $0.91$PFN–$0.09$PbTiO$_3$ (PFN–PT) ceramic samples prepared by original technology. This enabled performing measurements of dielectric properties and ME response up to the temperatures of 400-450 K without marked influence of conductivity. Obviously, the large conductivity of single crystals used in the previous studies [8, 9, 10] electrically shunts the sample already at T > 200 K and makes impossible correct measurement of the dielectric and PME response in single crystals at high temperatures. We also present a phenomenological theory of the PME effect, including its temperature dependence, which qualitatively describes well the measured data in both PFN and PFN-PT solid solution. In particular, it was found that the PME coefficient is proportional to the product of piezoelectric coefficient and the square of magnetic susceptibility.

**II.      Experimental**

Ceramic samples of PFN and 0.91PFN–0.09PbTiO$_3$ solid solution have been obtained by solid-state reaction route using high-purity Fe$_2$O$_3$, Nb$_2$O$_5$, PbO, and TiO$_2$. These oxides were batched in stoichiometric proportions, and 1 wt. % Li$_2$CO$_3$ was added to the batch. This addition promotes formation of the perovskite modification of PFN and reduces its conductivity [18]. After mixing thoroughly in an agate mortar under ethyl alcohol and subsequent drying, the green ceramic samples were pressed at 100 MPa in the form of disks of 10 mm in diameter and of 2-4 mm in height using polyvinyl alcohol as a binder. The sintering was performed at 1030-1070 $^0$C for 2 hours in a closed alumina crucible. The density of the obtained ceramics was about 92-97 % of theoretical one. The electrodes for measurements were deposited by silver paint (SPI Supplies, USA). Typical size of samples for measurements was 2.5x5x0.9 mm$^3$.

Dielectric measurements (capacitance and loss factor) were carried out in the 20 to 10$^5$ Hz range at temperatures 5-450 K using a HIOKI 3532-50 LCR HiTester. The piezoelectric coefficient, $d_{31}$, was measured using the standard resonance-antiresonance method. The polarization versus electric field (P-E) hysteresis loops were observed utilizing Sawyer-Tower experimental set up with programmable sweep of electric field. The ME coefficient was determined by a dynamic method [19] as a function of bias magnetic field $H_{dc}$ at small ac field $h_{ac}$ = 1-5 Oe and frequencies 0.2-1 kHz by measuring the voltage or current across the sample utilizing a lock-in-amplifier with high impedance preamplifier. High homogenous *ac* and *dc* magnetic fields were provided by conventional EPR spectrometer. Both fields were applied normal to the surface of the sample with electrodes. The sample was glued by silver past to metallic non-magnetic plate with the size 3x150 mm$^2$ and then connected to coaxial cable. The sample package was well screened in the magnetic field. Before the measurements, the sample was poled at room temperature by applying a *dc* electric field of 10 kV/cm for 30 min. In every experiment, more than two runs were repeated with the direction of $H_{dc}$ reversed and the change of the sign of the signal was confirmed. In this way, a possible spurious signal was distinguished from a true ME one whose sign is dependent on the *PH* product.

In our experiment, the PME effect is manifested as a polarization $P_{ac}$ induced by a small ac magnetic field $h_{ac}$ under application of dc field $H_{dc}$ [6, 9]. In a general approach, the magnetic field-induced components of the polarization can be obtained from the following free energy expansion [12]:

$$F(\bar{E},\bar{H}) = F_0 - P_i^s E_i - M_i^s H_i - \frac{1}{2}\varepsilon_0 \varepsilon_{ij} E_i E_j - \frac{1}{2}\mu_0 \mu_{ij} H_i H_j - \alpha_{ij} E_i H_j - \frac{1}{2}\beta_{ijk} E_i H_j H_k - ...$$

$$P_i = -\frac{\partial F}{\partial E_i} = P_i^s + \varepsilon_0 \varepsilon_{ij} E_j + \alpha_{ij} H_j + \frac{1}{2}\beta_{ijk} H_j H_k + ...$$

(1a)

where $P^s$ is the spontaneous polarization; $\alpha_{ij}$ and $\beta_{ijk}$ are linear and linear-quadratic ME coupling coefficients, respectively. With using collinear *dc* and *ac* magnetic fields $H = H_{dc} + h_{ac}\sin\omega t$, the first harmonic of the *ac* polarization detected by lock-in detector is:

$$P_{ac}(T) = \beta(T)H_{dc}h_{ac}. \qquad (1b)$$

Here $\beta(T)$ is the coefficient characterizing the studied PME effect in ceramic sample and at temperature *T*. Technically, this effect is described by a third rank tensor $\beta_{ijk}$ for the arbitrary orientations of polarization and magnetic field vectors. However, only an effective coupling constant, which represents an average of the different elements of the $\beta_{ijk}$ tensor, should be considered in ceramics. Note that a microscopic theory of the PME effect in the case of $C_{3h}$ symmetry was presented in Ref. 20.

**III. Experimental results**

The ME response was studied by measuring either the voltage or current across the sample. More precisely, the ME current is determined from Eq. (1b) as $I_{ME} = \dfrac{d(\beta S H_{dc} h_{ac} \sin\omega t)}{dt} = \beta\omega S H_{dc} h_{ac}$ with the lock-in (phase) detection at the frequency ω. Here S is the area of the sample. Consequently, the ME voltage is also proportional to both *dc* and *ac* magnetic fields via the simple expression $U_{ME} = I_{ME}(\omega C + 1/R_i)^{-1} \approx \dfrac{\beta H_{dc} h_{ac} S}{C}$, where C is the sample capacity. The expression for $U_{ME}$ is valid at the condition $(\omega C)^{-1} \ll R_i$, where $R_i \sim 10^9$ Ohm is the impendence of lock-in-amplifier with preamplifier. This relation was always fulfilled at the frequencies 0.2-1 kHz due to high capacities of the samples.

Figure 1 reports the room temperature ME voltage as a function of the applied *dc* magnetic field in a PFN ceramic, at two frequencies and $h_{ac} = 1.7$ Oe. Except for low values of the *dc* field, where the measured voltage is influenced by the background signal originating from a parasitic ferromagnetic phase, the ME voltage linearly increases with the strength of the *dc* magnetic field. The slope of the curves can be used to extract the coefficient β via the abovementioned formulas. This results in $\beta \approx 7.2\times10^{-19}$ s/A, which is about two orders of magnitude smaller than that the measured value in single crystal at 4-50 K [8-10]. As shown in Fig. 2, data were also obtained from measurements of the ME current.

It should be noted the ME response essentially increases in magnitude at the frequencies corresponding to the mechanical resonance of the sample holder, as illustrated in Figure 3, where the frequency dependence of ME current is presented for three values of *dc* magnetic field. At the mechanical resonance, the detected current increases by about 50 times.

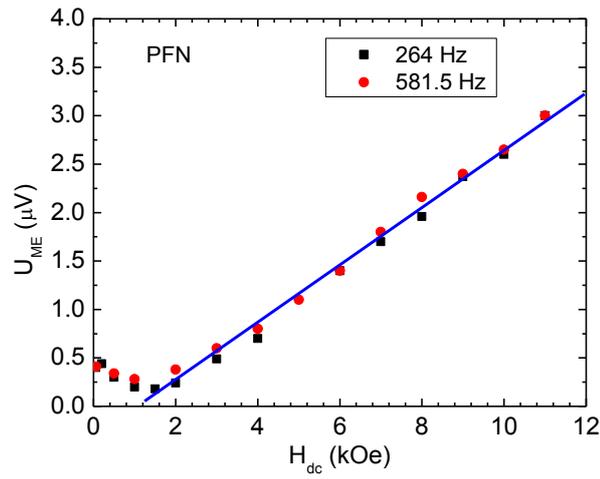

**Fig. 1**. Room temperature ME voltage in PFN as a function of applied dc magnetic field at two frequencies of *ac* field.

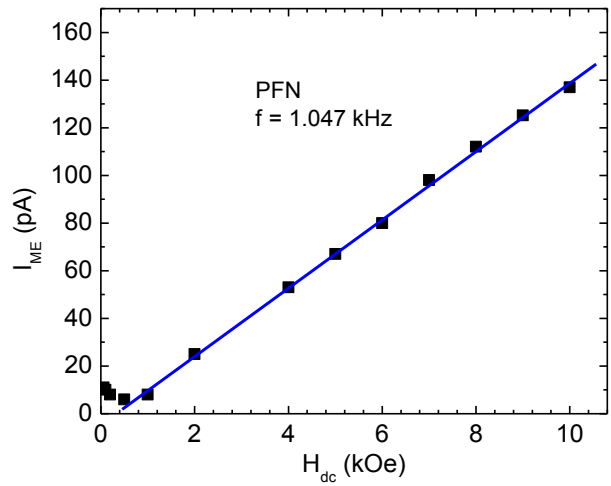

**Fig. 2**. Room temperature ME current in PFN as a function of applied *dc* magnetic field.

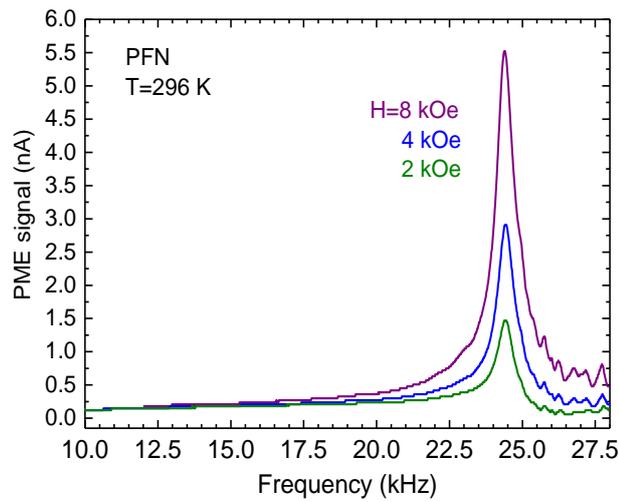

**Fig. 3**. Room temperature frequency dependence of the ME current in PFN measured for a *dc* magnetic field strength of 2, 4 and 8 kOe. It shows amplification of the PME signal at the mechanical resonance of the sample holder.

The PME effect requires a non-centrosymmetric lattice, in order to be finite. Therefore, it exists in the ferroelectric phase, but should vanish in the paraelectric one. To check such fact, we measured the temperature dependence of the PME coefficient for temperatures associated with the paraelectric phase. Figure 4 shows such data in a PFN ceramic sample by means of blue squares. It is seen that the PME coefficient β does not vanish at T > $T_N$ ≈ 150 K, unlike reported in Ref. 10 for single crystal. It rather increases up to the temperature of the paraelectric-to-ferroelectric phase transition and then rapidly decreases to zero in the paraelectric phase. Interestingly, this behavior is found to strongly correlate with the temperature dependence of the piezoelectric coefficient $d_{31}$ (open triangles in Fig. 4).

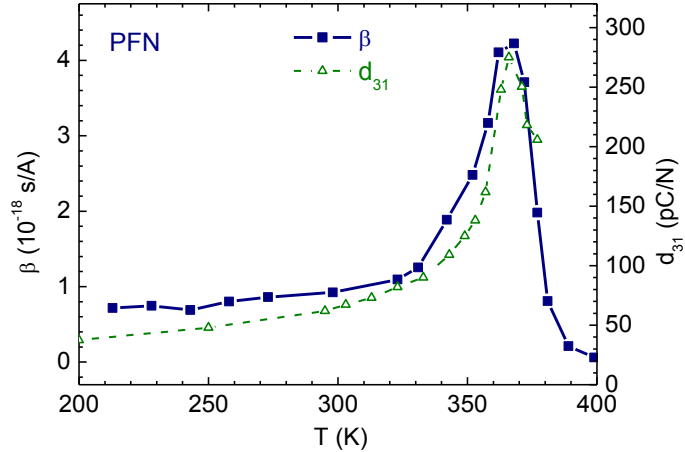

**Fig. 4**. Temperature dependence of the PME coefficient for PFN ceramics (solid line, filled squares). Dash line with open triangles represent the measured piezoelectric coefficients $d_{31}$.

The addition of PbTiO$_3$ to PFN, in order to form PFN-PT solid solutions, increases the temperature of the paraelectric-to-ferroelectric phase transition and the electric polarization at any temperature below $T_C$ of PFN, while the magnetic susceptibility practically does not change [21, 22]. The temperature dependence of the β PME coefficient measured in 0.91PFN–0.09PT ceramics is presented in Fig. 5, altogether with the data of the dielectric permittivity. One can see that the temperature dependence of the PME coefficient is similar to the temperature dependence of the dielectric permittivity up to the temperature of the paraelectric-to-ferroelectric phase transition, $T_C$ ≈ 397 K, above which the PME coefficient decreases abruptly to a noise level.

The main experimental results of the current work can thus be summarized as follows: (i) β increases upon adding PT and/or raising the temperature to the paraelectric-to-ferroelectric phase transition; (ii) the PME effect disappears in the paraelectric phase (but not in the paramagnetic ferroelectric phase); and (iii) its temperature dependence correlates well with the ones of the piezoelectric coefficient or dielectric permittivity in the ferroelectric paramagnetic phase.

Finally, let us compare the PME coefficient β values measured by us in PFN ceramics with those reported previously. Note that previous measurements were carried out at low temperatures, 5 K [8, 9] and 5-200 K [10] for single crystals. The values of $\beta_{333}(18K) \approx 1.0 \times 10^{-16}$ s/A [10] and $\beta_{333}(15K) \approx 1 \times 10^{-17}$ s/A [8] were reported, which are larger by two-three orders of magnitude from the one presently measured at room temperature. However, this is an expected result when taking into account that the PME coefficient in ceramics only represents an effective coupling constant (average of third rank tensor $\beta_{ijk}$). We should also emphasize that the ME response can acquire an additional increase in the magnetically ordered phase below the Neel temperature due to the interaction of magnetic and ferroelectric order parameters.

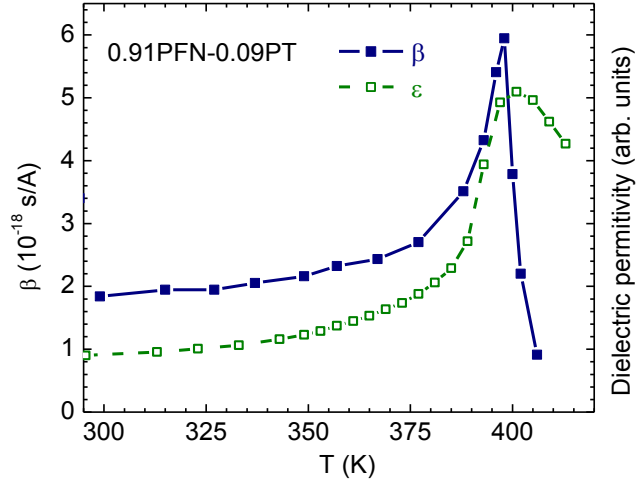

**Fig. 5**. Temperature dependence of the PME coefficient in 0.91PFN–0.09PT ceramics (solid line, filled squares). Dash line with open squares displays the dielectric permittivity in arbitrary units.

### IV. Phenomenological theory of paramagnetoelectric effect and its temperature dependence

Let us consider the phenomenological model of PME effect temperature dependence. The density of phenomenological free energy for the case of a bulk ferroelectric-antiferromagnet with ME coupling is the sum of polarization, magnetic and biquadratic magnetoelectric contributions. Nonzero terms in the free energy expansion are defined by the irreproducible representations of the material parent phase [14, 15, 16]. In this particular case:

$$G_{bulk} = G_P + G_M + G_{ME}, \tag{2a}$$

$$G_P = \frac{\alpha_P(T)}{2}P^2 + \frac{\beta_P}{4}P^4 + \frac{\gamma_P}{6}P^6 - PE, \tag{2b}$$

$$G_M = \frac{\alpha_L(T)}{2}L^2 + \frac{\beta_L}{4}L^4 + \frac{\alpha_M(T)}{2}M^2 + \frac{\beta_M}{4}M^4 - \mu_0 M H + \frac{\xi_{LM}}{2}L^2 M^2, \tag{2c}$$

$$G_{ME} = \frac{1}{2}\left(\xi_{MP}M^2 + \xi_{LP}L^2\right)P^2. \tag{2d}$$

Here $L$ is the antiferromagnetic order parameter, while $E$ and $H$ are the external electric and magnetic fields respectively. $\xi_{LM}$, $\xi_{MP}$ and $\xi_{LP}$ are biquadratic ME coefficients that couple corresponding order parameters. It should be noted, that only the coefficients in front of $P^2$, $L^2$ and $M^2$ are assumed to be dependent on the temperature, namely $\alpha_P(T) = \alpha_P^{(T)}(T - T_C)$, $\alpha_L(T) = \alpha_M^{(T)}(T - T_N)$ and $\alpha_M(T) = \alpha_M^{(T)}(T-\theta)$ where $T_C$ is the temperature of the paraelectric-to-ferroelectric phase transition, $\theta$ is the magnetic Curie temperature, and $T_N$ is the Neel temperature. Hereinafter, we regard that the magnetic phase transitions are of the second order, while the paraelectric-to-ferroelectric phase transitions can be of the second order (II-nd order PT) if $\beta_P > 0$ and $\gamma_P \geq 0$, or of the first order (I-st order PT) if $\beta_P < 0$ and $\gamma_P > 0$.

In the case of PFN, the ferroelectric phase transition takes place at 360-370 K. The transition to AFM phase occurs at the Neel temperature $T_N \approx 150$ K, and the magnetic Curie temperature $\theta$ is negative for PFN, that is about $-520$ K (see, e.g. Refs. 23, 24).

For the I-st order PT, the polarization can be obtained from the minima of the free energy (2) as follows:

$$P(T) = \sqrt{\frac{\sqrt{\beta_P^2 - 4(\alpha_P(T) + \xi_{MP}M^2 + \xi_{LP}L^2)\gamma_P} - \beta_P}{2\gamma_P}} \quad (3a)$$

For the II-nd order PT with $\gamma_P = 0$ the expression is simplified to $P(T) = \sqrt{-(\alpha_P(T) + \xi_{MP}M^2 + \xi_{LP}L^2)/\beta_P}$. Assuming that the magnetization M is linearly proportional to the applied magnetic field, $M \approx \chi_{FM}(T)H$, and the expression (3a) could be formally expanded as a series on $M$:

$$P(T) \approx P_S(T)(1 - \chi_{FE}(T)\xi_{MP}M^2) \approx P_S(T)(1 - \chi_{FE}(T)\xi_{MP}(\chi_M(T)H)^2) \quad (3b)$$

The spontaneous polarization $P_S(T)$ is given by the formulae (3a) at $M=0$. Dielectric susceptibilities in the ferroelectric phase and magnetic susceptibility, $\chi_{FE}(T)$ and $\chi_M(T)$, are given by the following expressions:

$$\chi_{FE}(T) = \frac{-1}{4(\alpha_P(T) + \xi_{LP}L^2) - (\beta_P^2/\gamma_P)\left(1 + \sqrt{1 - 4(\beta_P^2/\gamma_P)(\alpha_P(T) + \xi_{LP}L^2)}\right)} \quad (4a)$$

and

$$\chi_M(T) = \frac{\mu_0}{\alpha_M^{(T)}(T-\theta) + \xi_{LM}L^2 + \xi_{MP}P_S^2(T)}. \quad (4b)$$

Note that, for the II-nd order PT with, $\gamma_P = 0$ Eq. (4a) can be simplified as $\chi_{FE}(T) = -1/2(\alpha_P(T) + \xi_{LP}L^2)$, and that Eq. (4b) strongly correlates with analytical results obtained in Refs. 14, 25. Equations (4a) and (4b) are valid in the ferroelectric-antiferromagnetic phase ($L \neq 0$) but also in the ferroelectric – paramagnetic phase (i.e. at $M = L = 0$).

From the comparison of Eq. (1b) with Eq. (3b) one can thus derive the following equation for the coefficient β characterizing the PME effect:

$$\beta(T) = -P_S(T)\chi_{FE}(T)(\chi_M(T))^2 \xi_{MP} \quad (5)$$

Interestingly, such latter formula is also consistent with the one previously derived in Ref. 25. Note that the temperature dependence of the product $P_S(T)\chi_{FE}(T)$ completely determines the temperature dependence of the corresponding component of the piezoelectric coefficient $d_{31}(T)$ in the thermodynamic limit. Besides, the strong inequality $|\xi_{MP}M^2| \ll |\alpha_P(T)|$, formally required for the validity of the expansion (3b), is certainly invalid in the immediate vicinity of the II-nd order PT, where $\alpha_P(T)\to 0$. Equation (5) is the main result of the phenomenological model, implying that the temperature dependence of the PME coupling coefficient is determined by the product of polarization, dielectric susceptibility and the square of magnetic susceptibility with the temperature dependences of some of these quantities being given by Eqs. (4).

Allowing for the small influence of the biquadratic coupling on magnetic susceptibility (4b) and considering the paramagnetic phase (i.e., at $L=0$) for the purpose of illustration, one could get from Eq. (5) that the temperature dependence of PME effect coefficient has the following form in this case:

$$\beta(T) \propto \begin{cases} \dfrac{-\zeta\left(T - \theta + \zeta\left(\sqrt{1-\lambda(T/T_C - 1)} \pm 1\right)\right)^{-2} \sqrt{\sqrt{1-\lambda(T/T_C - 1)} \pm 1}}{\lambda(T/T_C - 1) - \left(1 \pm \sqrt{1-\lambda(T/T_C - 1)}\right)}, & T < T_C, \\ 0, & T > T_C. \end{cases} \qquad (6)$$

The sign "+" in expression (6) corresponds to the I-st order PT, and "−" to the II-nd order PT. Expression (6) depends only on the ratio $\eta = \theta/T_C$, effective ME parameter $\zeta = \xi_{MP}\dfrac{\alpha_P^{(T)}\beta_P}{\alpha_M^{(T)}2\gamma_P}$ and on the dimensionless parameter $\lambda = \dfrac{4\gamma_P}{\beta_P^2}\alpha_P^{(T)}T_C$. For the II-nd order PT with $\gamma_P = 0$, the expression is simplified as $\beta(T) \propto \dfrac{\xi(T - \theta + \xi(T_C - T))^{-2}}{\sqrt{T_C - T}}$, where the parameter $\xi = \xi_{MP}\dfrac{\alpha_P^{(T)}}{\alpha_M^{(T)}\beta_P}$.

For PFN, the ratio $\eta$ is close to $-1.4$, because $\theta = -520$ K and $T_C = 370$ K. The parameter $\lambda$ is equal to 42 for PbTiO$_3$ and typically is not less than 10 for other ferroelectrics, including PFN. The parameters $\xi$ and $\zeta$ are unknown, because the biquadratic coupling constant $\xi_{MP}$ value and its temperature dependence are unknown. Such latter coupling constant can be of arbitrary sign and is typically small. For instance, we did not find any visible change of the dielectric permittivity and piezoresonance frequency at the AFM phase transition (Fig. 6), indicating that the biquadratic coupling is very small as compared, for instance, with EuTiO$_3$ where the dielectric permittivity decreases abruptly by almost 5% on cooling below the Neel temperature [26]. Note that both dielectric permittivity and piezoresonance frequency were measured with the accuracy of $10^{-4}$. Similar results were obtained for both ceramics and single crystal of PFN. A value of $\xi_{MP}$ of $-0.00503$ J m$^3$/(C$^2$A$^2$) was estimated from the fitting of the experimental results on magnetic susceptibility measured in this work by a Curie-Weiss law with biquadratic ME coupling term included (see Eq. (4b)). Note also that our data shown in Fig. 6 do not support the results obtained in highly conductive PFN single crystal [27] where huge step-like drop of the dielectric permittivity below the Neel temperature from about 250 to few units was reported.

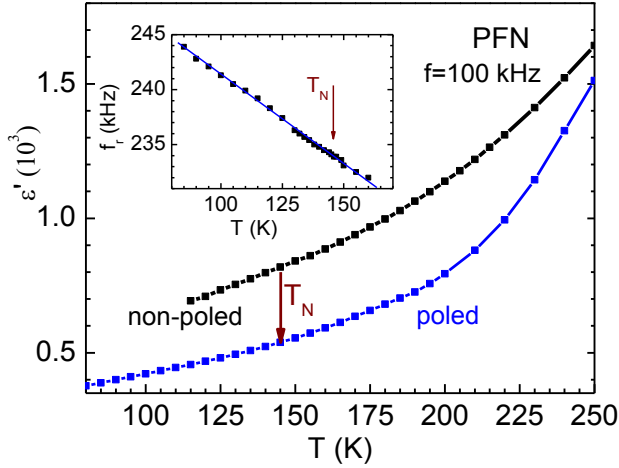

**Fig. 6**. Temperature dependence of the dielectric permittivity measured for poled and non-poled PFN ceramics. The inset shows the temperature dependence of the piezoresonance frequency. Both dielectric permittivity and piezoresonance frequency have no visible changes at the AFM phase transition.

Examples of the PME coefficient β dependences on temperature, as predicted by Eq. (6), are shown in Figs. 7(a-c) for the case of paramagnetic ferroelectrics with different values of θ in units of $T_C$ and different λ parameter. The increase of the PME coefficient at low temperatures is caused by the enhancement of the magnetic permittivity on cooling, and its divergence at $T_C$ is caused by the divergence of the dielectric permittivity at the paraelectric-to-ferroelectric phase transition (see Fig. 7(d)).

The influence of the negative ratio η is qualitatively the same for both I-st and II-nd order PTs, namely a decrease of its absolute value leads to a strong increase of β when the temperature decreases. Quantitatively, this increase is much less pronounced for the II-nd order PT than that for the I-st order one (compare different curves in Figs. 7(a) and 7(b)). The increase of λ at fixed η leads to the monotonic increase of the PME coefficient β (compare different curves in Fig. 7(c)).

## V. Discussion

As indicated above, the presently developed phenomenological Landau theory predicts that the PME coefficient is simply the product of the piezoelectric coefficient $d_{31}$, square of magnetic susceptibility and the coefficient of biquadratic magnetoelectric coupling, $\beta(T) = -d_{31}(T)(\chi_M(T))^2 \xi_{MP}$. As a result, above room temperature, the temperature dependence of the PME coefficient in PFN is mainly determined by the piezoelectric coefficient $d_{31}$ (that shows critical behavior near the temperature of the paraelectric-to-ferroelectric phase transition, and which is also proportional to the dielectric response), since the magnetic susceptibility is rather insensitive to temperature there. Such prediction is fully consistent with the good correlation found between the experimental measured temperature dependences of the PME coefficients and piezoelectric coefficient $d_{31}$ (see Fig. 4) or dielectric permittivity (Fig. 5).

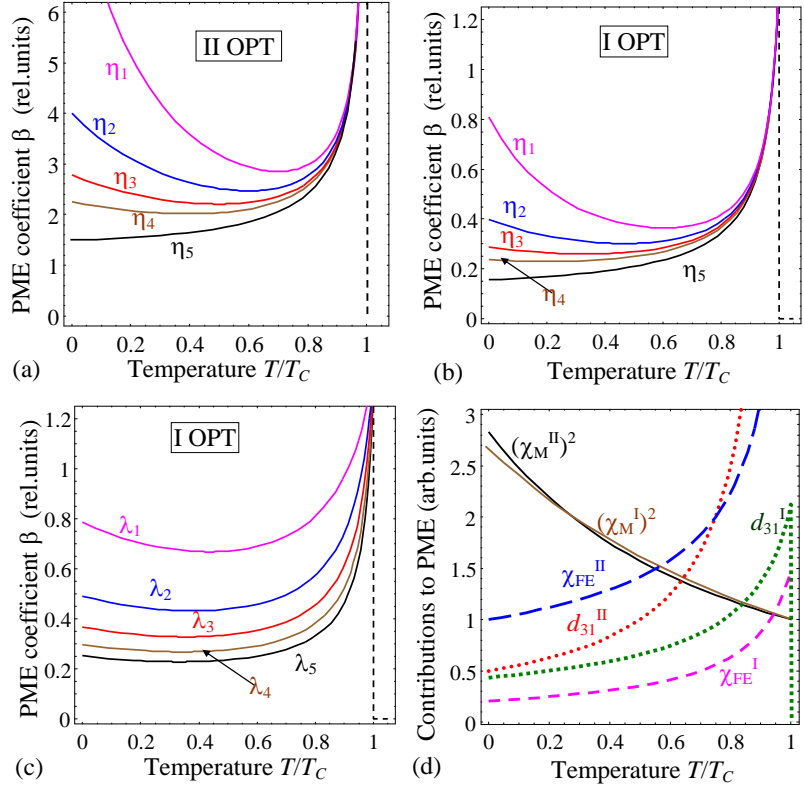

**Fig. 7**. Temperature dependence of the PME coefficient β calculated from Eq. 6 for ferroelectrics having a II-nd order PT with $\gamma_P = 0$ (a) and I-st order PT (b-c). The parameters λ = 42, ξ = –0.01, η = –0.5, -1, -1.5, -2 and -5 (shown near the curves as $\eta_1, \ldots, \eta_5$) in the plots (a-b). For the plot (c) η = –1.44, ζ = –0.01, λ = 10, 20, 30, 40 and 50 (shown at the curves as $\lambda_1, \ldots, \lambda_5$). (d) Temperature dependences of the square of the magnetic susceptibility ($\chi_M^2$), dielectric susceptibility ($\chi_{FE}$), and piezoelectric coefficient ($d_{31}$) calculated from Eq. (6) for ferroelectrics having a I-st or II-nd order PT.

The theoretical dependences of the PME coefficient as a function of different parameters are illustrated in Figs. 7(a-c). In particular, one can see that the PME coefficient increases again at low temperatures (for T<<T$_C$), which is essentially due to an increase of the magnetic susceptibility. Such tendency was indeed observed experimentally in PFN single crystal [10].

Furthermore, we should mention the results of Ref. 28, where the ME coefficient, defined as $\alpha_{ME} = \dfrac{U_{ME}}{h_{ac} d}$ (d is thickness of sample), was measured in PFN ceramics at room temperature in the paramagnetic phase. Surprisingly, this ME coefficient almost linearly increased when increasing the *ac* field frequency and did not depend on the bias *dc* magnetic field. These features, altogether with the unrealistic large $\alpha_{ME} = 20-40$ mV/(Oe cm), strongly suggest that the measured voltage in Ref. 28 was of inductive origin, as further emphasized by the fact that the ME coefficient in PFN calculated from our data in similar units (even under bias field of 10 kOe) is only 18 μV/(Oe cm) for the sample with area 0.11 cm$^2$.

Finally, let us also indicate that this the present work opens doors for other related studies. For instance, it is striking to realize that (i) the PME effect measured here is in fact technically an *ac* effect, as consistent with Eq. (1b) and Fig. 3, since the total applied field contains both *dc* and ac components, and (ii) that we are not aware of any previous atomistic simulation that has ever reported the computation of *ac* PME coefficients, which contrasts with the case of the static (*dc*) PME [29]. One may thus wonder if it is technically feasible to compute such *ac* coefficient at an atomistic level, which may result (in future studies) in the discovery of an enhancement of such coefficient by playing with the frequency of the applied *ac* field.

To tackle this interesting problem, we decided to perform simulations on the well-known $BiFeO_3$ (BFO) magnetoelectric bulk material. This material was selected for several reasons. First of all and unlike for PFN and PFN-PT, effective Hamiltonians are available for it [29, 30, 31]. Secondly, it bears similarity with PFN and PFN-PT, in the sense that the paraelectric-to-ferroelectric transition occurs at a much higher temperature than the magnetic Neel temperature. Thirdly and as evidenced in Refs. 29-30, its linear ME effect can be safely neglected in front of its PME effect for large enough magnitude of the applied magnetic field. As a result, applying a (large) magnetic field of the form $H = H_{dc} + h_{ac}\sin\omega t$ should result in BFO bulks in a time-dependent polarization given by:

$$P(t) = P(t=0) + \beta H_{dc} h_{ac} \sin(\omega t) + \tfrac{1}{2}\beta h_{ac}^2 \sin^2(\omega t) \tag{7}$$

Such Equation therefore tells us that the coefficients in front of $\sin(\omega t)$ and $\sin^2(\omega t)$ should both be related to $\beta$.

Here, we implement the effective Hamiltonian provided in Ref. 31 within a molecular dynamics (MD) scheme that treats on the same footing the dynamics of structural quantities (such as the electrical dipole moments) and the dynamics of the magnetic moments [32], in order to determine if atomistic simulations can reproduce Eq. (7) and accurately predict the associated *ac* PME coefficient. Practically, we apply very large *dc* and *ac* fields, of 490 T and 245 T magnitudes, respectively, along the [11-2] pseudo-cubic direction, and follow the time dependence of the electrical polarization along its pseudo-cubic [111] direction. Note that choosing such large magnitude of the magnetic fields is necessary in order to hope to accurately extract the $\beta$ coefficient since this latter is typically rather small (selecting small magnetic fields would make the polarization fluctuating at the same level than the computational noise, which would thus make an accurate determination of $\beta$ impractical). The frequency *f* of the applied *ac* magnetic field is chosen to be 5 GHz (correspondingly, the angular frequency ω is $31.4\times10^9$ rad/s), and the calculations are conducted at 1 K.

Figures 8 reports the resulting time dependency of the polarization, as well as its fitting by Eq. (7). One can see that the polarization has some fluctuations (especially around its extrema), but does tend to obey Eq. (7) in overall. Interestingly, the $\beta$ parameter obtained from the fitting of the MD data by Eq. (7) is found to be equal to (i) $-4.65\times10^{-8}$ $C/m^2T^2$ = $-0.73\times10^{-19}$ s/A, when concentrating on the coefficient in front of $\sin(\omega t)$ in Eq. (7); and (ii) of $-4.72\times10^{-8}$ $C/m^2T^2$ = $-0.75\times10^{-19}$ s/A, when considering the coefficient in front of $\sin^2(\omega t)$ in Eq. (7).

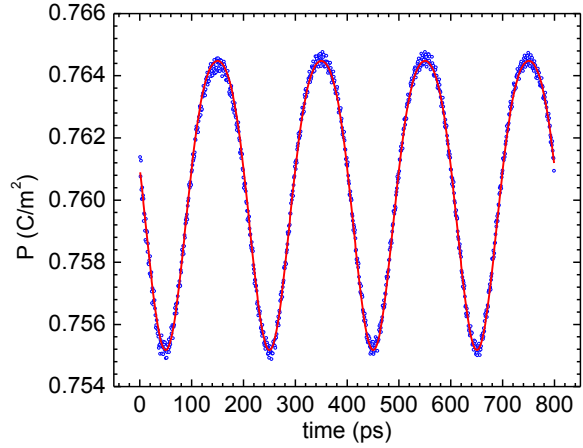

**Fig. 8**. Time dependency of the polarization of BiFeO$_3$ bulk under a magnetic field $H = H_{dc} + h_{ac} \sin \omega t$, with H$_{dc}$ = 490 T, h$_{ac}$ = 245 T and ω = 31.4×10$^9$ rad/s, as predicted by atomistic simulations combining MD simulations and an effective Hamiltonian scheme. The symbols show the MD data, while the solid line represents the fit of these data by Eq. (7).

The good agreement between these two (differently-obtained) $\beta$ parameters, as well as Fig. 8, demonstrates that our present atomistic simulations can indeed mimic the *ac* PME effect. Such fact is further emphasized when realizing that our predictions are rather similar to the corresponding magnitude of 0.6x10$^{-19}$ s/A that was measured in Ref. 33 for $\beta_{311}$. Note that we also calculated the static (*dc*) PME effect in bulk BFO (not shown here) via the combination of the aforementioned effective Hamiltonian [31] and a Monte-Carlo technique, by solely applying a *dc* magnetic field and computing the dependency of the polarization with the magnitude of this *dc* magnetic field – as similar to what was done in Refs. 29-30. We numerically found that the resulting static $\beta$ is about $-3.69\times10^{-8}$ C/m$^2$T$^2$ = $-0.58\times10^{-19}$ s/A, i.e. it is slightly smaller in magnitude than the computed *ac* PME coefficients (note that the computed $\beta$ coefficients of BFO are smaller by about one order of magnitude than those presently measured in PFN ceramics (cf Fig. 4) likely because (1) the predictions are made at 1 K while the measurements are done at much higher temperature; (2) the Curie temperature of BFO is about three times larger than that of PFN; and (3) magnetoelectric coefficients can, of course, vary in magnitude from one system to another). It will thus be interesting to check in future studies if one can further and even more dramatically enhance the *ac* PME coefficients by applying an *ac* magnetic field for which the frequency is close to the resonant one of phonons or magnons (rather than 5 GHz as done here), especially since a previous phenomenological study predicted a large increase of the *linear* dynamical ME effect for these resonant frequencies [34]. Studying other types of dynamical magnetoelectric effects, such as those predicted in Ref. [35], can also be of great technological and fundamental interest [35].

## V. Conclusion

We have experimentally found an anomalous temperature behavior of the paramagnetoelectric coefficient in high-resistive ceramics of multiferroic PFN and in PFN-based solid solution 0.91PFN-0.09PbTiO$_3$. The value of the PME coefficient shows a pronounced maximum near the ferroelectric phase transition temperature and then decreases quite sharply down to zero on a subsequent temperature increase. However, the PME coefficient is sufficiently small, $(1 – 6) \times 10^{-18}$ s/A, as compared to values measured in PFN single crystals at temperatures 15-40 K where it reaches values of $10^{-17}–10^{-16}$ s/A [8,10]. In PFN-PT solid solution, the temperature interval of the PME effect is extended to higher temperatures in accordance with the shift of the paraelectric-to-ferroelectric phase transition temperature with adding PbTiO$_3$.

The theoretical description of the PME effect, described in the framework of the Landau theory of phase transitions, describes rather well the temperature behavior of the PME coefficient. In particular, the Landau theory predicts that the PME coefficient is equal to the product of the spontaneous polarization, dielectric permittivity, square of magnetic susceptibility and the coefficient of biquadratic ME coupling. A qualitative agreement between the measured temperature dependence of the PME coefficient and that theoretically predicted is found. We also discussed and tested the possibility of using atomistic simulations in order to determine if the *ac* PME effect can be enhanced by varying the frequency of the *ac* applied magnetic field.

## Acknowledgements

The research was supported by the GA CR under project No. 13-11473S, the Russian Foundation for Basic Research (project 14-02-90438_Ukr_a), National Academy of Sciences of Ukraine (Grant No. 07-02-14) and the Ministry of education and science of Russian Federation (research project 2132). S.P. and L.B. acknowledge the Air Force Office of Scientific Research under Grant FA9550-16-1-0065.